\documentstyle[12pt]{ioplppt}    
\begin{document}
\author{J. Phys. A: Math. Gen.
32 (1999) 563-568. Printed in the UK}

\paper{Semiclassical treatment of the logarithmic
perturbation theory
}[ Semiclassical
 treatment of the logarithmic
perturbation theory]


\author{I V Dobrovolska and R S Tutik\ftnote{3}{To
whom correspondence should be addressed.

E-mail address: tutik@ff.dsu.dp.ua}}

\address{Department of Physics, Dniepropetrovsk State
University, Dniepropetrovsk,
UA-320625, Ukraine
}

\address{Receved 23 June 1998, in final form 28 October 1998}
\maketitle

\begin{abstract}
The explicit semiclassical treatment of logarithmic
perturbation
theory for the nonrelativistic bound states problem is
developed. Based upon
$\hbar$-expansions and suitable quantization
conditions a new procedure for deriving
perturbation expansions
for the one-dimensional anharmonic oscillator is offered.
Avoiding
disadvantages of the standard approach, new handy recursion
formulae
with the same simple form both for ground and exited states
have been obtained.
As an example, the perturbation expansions for
the energy eigenvalues of the harmonic oscillator
perturbed by
 $\lambda x^{6}$ are considered.

\end{abstract}
%
%

\section{Introduction}
Logarithmic perturbation theory [1-8] 
is one of the principal approximation
techniques in theoretical and mathematical physics. 
Within the framework of this theory,
the conventional way to solve a
quantum-mechanical bound-state problem
consists in changing from the wave function 
to its logarithmic derivative 
and converting the time-independent Schr{\"o}dinger 
equation 
into the nonlinear Riccati equation.
Such a procedure 
leads to handy recursion relations in the case of ground states, 
but becomes extremely cumbersome in the description of radial 
excitations
when nodes of wave function are  taken into account.
Although several attempts have been made to improve the method in
the latter case [9-11], they have not resulted 
in a desirable simple algorithm.

On the other hand, it is well known, 
that the radial quantum number, $n$,
most conveniently and naturally is introduced 
into consideration by means of quantization conditions,
as  in the Wentzel-Kramers-Brillouin (WKB) approximation [12-14].
However, the WKB approach is more suitable
for obtaining energy eigenvalues in the limiting case
of large quantum numbers,
whereas the perturbation theory deals with low-lying levels.
Usually, the perturbation results are obtained within the framework
of the WKB method by recasting the WKB expansions [15-19].

The objective of this paper 
is to develop an explicit semiclassical treatment
of logarithmic perturbation theory
and to describe a straightforward semiclassical procedure 
for obtaining the perturbation corrections 
through handy recursion formulae,
having the same form
both for ground and excited states.

For the sake of simplicity, 
we restrict ourselves
to the consideration of the bound-state problem 
for the one-dimensional anharmonic oscillator. 
The generalization to the three-dimensional
problem reqiures including the quantization conditions
for orbital momentum 
and will be published elsewhere.

\section{ Method}
The system to be treated 
is described by the Schr{\"o}dinger equation

\begin{equation}
-{{\hbar}^2\over {2m}}U''(x)+V(x)U(x)=EU(x)
\end{equation}
where the potential function,  $V(x)$  , 
has a simple minimum 
and hence can be given by the expression

\begin{equation}
V(x)=\frac{1}{2}m\omega^{2}x^{2}+\sum_{i{\geq}1}f_{i}x^{i+2}  .
\end{equation}

With changing the scale of the variable, $x\to\sqrt{\hbar}x$ ,
it becomes obvious 
that the coupling constants, $f_i$ , 
appear in common with powers of Planck's constant, $\hbar$ .
Therefore, the perturbation series 
must be semiclassical  $\hbar$-expansions, too.
Although well known in the folk
wisdom of theoretical physics, this assertion was,
nevertheless, proved not so long ago [20]. 
It has been argued 
that the energy eigenvalues 
under consideration should be concentrated near the minimum of
the potential and
should behaved as

\begin{equation}
E=\hbar\omega (n+\case12)+\sum_{i\geq2}E_{i}(\omega,n)\hbar^{i}.
\end{equation}

To our knowledge, 
there are only a few procedures 
for computing the coefficients  $E_{i}(\omega,n)$. 
They involve, in particular, applying the methods 
of the comparison equation [21] 
and complex "sprout" [22]; 
an analytic continuation in the  $\hbar$ -plane [23]; 
various approaches within the framework of the 
WKB approximation [15-19]; 
 quantization using the methods of classical mechanics [17,24]; 
and, lastly, expansions in the  $\hbar^{1/2}$  -series [25]. 
However, all of these methods have some disadvantages.

Here we propose a new, simpler and more straightforward
semiclassical  technique.
Being explicitly opposed to the WKB approach,
 it is based on different quantization conditions
which are more appropriate for describing 
the solution of the bound-state problem 
in the vicinity of a potential minimum.

Following usual practice, we apply the substitution, $C(x)=\hbar
U'(x)/U(x)$ , accepted in the logarithmic perturbation theory and
go over
from the Schr{\"o}dinger equation (1) to the Riccati equation

\begin{equation}
\hbar C'(x)+C^{2}(x)=2m[V(x)- E].
\end{equation}
We attempt to solve it in a semiclassical manner with series
expansions in the Planck constant
\begin{equation}
E=\sum^{\infty}_{k=0}{E_{k}\hbar^{k}}\;\;\;\;\;
C(x)=\sum^{\infty}_{k=0}C_{k}(x)\hbar^{k}
\end{equation}
that result in the system
\begin{eqnarray}
C_{0}^{2}=2m[V(x)-E_{0}]\nonumber\\
C'_{0}+2C_{0}(x)C_{1}(x)=-2mE_{1}\nonumber\\
\cdots \\C'_{k-1}(x)+\sum_{i=0}^{k}C_{i}(x)C_{k-i}(x)=-
2mE_{k}.
\nonumber
\end{eqnarray}

In the case of ground states, this system coincides with one
derived by means
of the standard technique and can be solved
straightforwardly.
However, complications of the logarithmic perturbation theory
arise in the
description of radial excitations when the
nodes of wavefunctions are 
included in some separate factor. We intend
to circumvent
these difficulties by making use of the quantization conditions.
 The matter
of the latter consists in applying the principle of argument,
known from the
analysis of complex variables, to the logarithmic derivative, $C(x)$ .

Since the wavefunction of the $nth$ radially excited state
has $n$ real zeros we have
\begin{equation}
\frac{1}{2\pi\i}\oint{C(x)\,\d x}=
\frac{1}{2\pi\i}\sum^{\infty}_{k=0}{\hbar^{k}\oint{C_{k}(x
)\,\d x}}=\hbar n.
\end{equation}
There is, however, one important point to note. In the
WKB approach, this
condition is supplemented by the following rule of
achieving a classical limit:
\begin{equation}
\hbar\to 0\;\;\;\;n\to\infty\;\;\;\;\hbar n={\rm const}
\end{equation}
accompanied by the equality of the quantum energy to the
classical one.

In contrast to this, our method, 
dealing with low-lying states
and being complementary to the WKB approach,
involves the alternative possibility
 \begin{equation}
\hbar\to 0\;\;\;\;n={\rm const}\;\;\;\;\hbar n\to 0
\end{equation}
which was formerly applied in deriving coefficients of the $1/N$
expansions [26-29].
In the limiting case, as  $\hbar\to 0$, a particle is
now
lowered to the bottom of a potential well and its classical energy
becomes
$E_{cl}=minV(x)$ , which equals to zero in our case.

Thus, in view of the rule (9), the quantization conditions
(7) become
\begin{equation}
\frac{1}{2\pi\i}\oint{C_{1}(x)\,\d x}=n \;\;\;\;\;
\frac{1}{2\pi\i}\oint{C_{k}(x)\,\d x}=0  \;\;\;\; k>1.
\end{equation}

A further application of the theorem of residues to the explicit
form of
functions  $C_{k}(x)$ easily solves the problem of taking
into
account nodes of the wavefunctions.

\section{ Recursion  formulae}
Let us consider the system (6) and investigate the behaviour of the
functions  $C_{k}(x)$.
From the first equation it is seen that
\begin{equation}
C_{0}(x)=-{[2mV(x)]}^{1/2}
=-m\omega x({1+{2\over{m\omega^2}} \sum_{i\geq
1}{f_{i}x^i}})^{1/2}=
x\sum^{\infty}_{i=0}{C^{0}_{i}x^i}
\end{equation}
where the minus sign is chosen from boundary conditions,
and coefficients
 $C^{0}_{i}$  are defined by parameters of the potential as
\begin{equation}
C^{0}_{0}=-m\omega\;\;\;\; 
C^{0}_{i}={1\over{2m\omega}}
({\sum_{p=1}^{i-1}{C^{0}_{p} C^{0}_{i-p}-2mf_{i}}})
\;\;\;\;\;i\geq 1.
\end{equation}
Because the point $x=0$ is a simple zero for the
function  $C_{0}(x)$
, the function $C_{k}(x)$ has a pole of
the order of  $(2k-1)$ at this point and consequently can be
represented by a
Laurent series
\begin{equation}
C_{k}(x)= x^{1-
2k}\sum^{\infty}_{i=0}{C^{k}_{i}x^i}\;\;\;k\geq 1.
\end{equation}

Then, according to the theorem of residues, the
quantization conditions
(10), expressed in terms of the Laurent series coefficients, take an
especially simple
form
\begin{equation}
C^{k}_{2k-2}=n\delta_{1,k}
\end{equation}
where the symbol $\delta_{1,k}$ is the Kronecker delta.

Substituting the expansions (11) and (13) into (6) and
equating coefficients of
equal powers of $x$, we derive
\begin{equation}
(3-2k+i)C^{k-1}_{i }+\sum_{j=0}^{k}\sum_{p=0}^{i}
C^{j}_{p}C^{k-j}_{i-p}=
-2mE_{k}\delta_{i,2k-2}.
\end{equation}
Sorting out the case of  $i \not= 2k-2 $ yields the recursion
relation for obtaining the coefficients $C^{k}_{i}$ :
\begin{equation}
C^{k}_{i}=-{1\over{2C^{0}_{0}}}
[{(3-2k+i) C^{k-1}_{i }+\sum_{j=1}^{k-1}\sum_{p=0}^{i}
C^{j}_{p}C^{k-j}_{i-p}
+2\sum_{p=1}^{i}C^{0}_{p}C^{k}_{i-p}}]
\end{equation}
whereas putting $i=2k-2$ we would find the recursion formula
for the energy
eigenvalues
\begin{equation}
2mE_{k}=- C^{k-1}_{2k-2 }-
\sum_{j=0}^{k}\sum_{p=0}^{2k-2} C^{j}_{p}C^{k-j}_{2k-2-p}
\; .
\end{equation}

Thus, equations (16) and (17) determine coefficients of
perturbation expansions of energy eigenvalues and eigenfunctions in
the same simple form both for the ground and excited states.

It should be noted, that just conditions (9),
represented in the form (14)
simplify the consideration of the perturbation theory
by means of semiclassical expansions.
It becomes more evident under comparison of our technique 
with the rescaled version of the WKB approach [17-19],
where the order of the energy is taken into account 
by the equality $E= \hbar \varepsilon $. In this case, the coefficients
$C_k (\varepsilon , x)$ in equation (7) have poles, too. 
The WKB quantization condition (8) then reads as
\begin{equation}
\sum\limits_{k=0}^\infty  {\hbar ^{k-1}Res\:C_k\left( {\varepsilon ,x} \right)}=n
\end{equation}
with residues polynomial with respect to $\varepsilon$.
By truncating the series in the left-hand side we arrive at
the equation for determination of approximations to 
the energy eigenvalues $\varepsilon$, and only after
subsequent re-expanding, we can restore the results of
perturbation theory [17-19].

\section{ Discussion and examples} 
From equation (17) it is readily seen 
that for the energy eigenvalues, when $k=1$, we immediately have
the oscillator approximation
\begin{equation}
E_{1}=\omega(n+{\case12})
\end{equation} 
and with $k=2$ one
obtains the form familiar from standard textbooks [15] 
\begin{equation}E_{2}=- {\frac{15f^{2}_{1}}{4m^{3}\omega^{4}}}
(n^{2}+n+{\case{11}{30}})+
{{3f_{2}}\over{2m^{2}\omega^{2}}}( n^{2}+n+{\case12}).
\end{equation}

It is easy to demonstrate that in the case of the harmonic oscillator
our technique restores the exact solution for the wave functions
as well.

Putting, for simplicity, $\hbar=m=\omega=1$, from equations (12),
(13) and (16) we find
\begin{equation}
C_{0}(x)=-x\;\;\;\;\;\;C_{k}(x)=d_{k}x^{1-2k}\;\;\;\;k>0
\end{equation}
where
\begin{equation}
\eqalign{2d_{k}=(3-2k) d_{k-1} +\sum_{j=1}^{k-1}d_{j}d_{k-j}
\;\;\;\;k>1\\
d_{1}=n.}
\end{equation}
Carring out the integration of the function $C_{0}(x)$
we obtain the exponential factor of the eigenfunction.
Its remaining part is a polynomial that satisfies the
equation
\begin{equation}
P'_{n}/P_{n}=\sum_{k=1}^{\infty}d_{k}x^{1-2k}
\end{equation}
and, consequently, has a form
\begin{equation}
P_{n}(x)=x^{\sigma}\sum_{i=0}^{m_{0}}a_{i}x^{2i}
\;\;\;\;\sigma=0 \;\; or \;\; 1 \;\;\;\;\;
n=2m_{0}+\sigma .
\end{equation}
On the basis of equation (23), the polynomial coefficients
, $a_{i}$, are determined by the system
\begin{equation}
(n-2m-\sigma)a_{m}+d_{2}a_{m+1}+...+d_{m_{0}-m+1}a_{m_{0}}=0.
\end{equation}
The combination of these equations, multiplyed by a suitable $d_{j}$
with a view of taking into account equation (22), arrives at the
following
relation between two consecutive coefficients:
\begin{equation}
a_{m}=-a_{m+1}{(2m+\sigma+2)(2m+\sigma+1)\over4(m_{0}-m)}
\end{equation}
that is the recursion formula for the Hermite polynomials ( see, for
instance [30]).

And at last, as an example, we consider the anharmonic
oscillator with the potential
\begin{equation}
V(x)={\case12}x^{2}+{\case12}\lambda x^{6}.
\end{equation}
Though this oscillator is widely discussed in the scientific
literature [31-33], the analytical expressions for the perturbation
expansions of energy eigenvalues presented in [31] are
incorrect. The correct expansion coefficients obtained
by means of formulae (16) and (17) have the form
\begin{equation}
\eqalign{
E_{1}=n+{\case12}\\
E_{3}=5 \, \lambda
\, 2^{-4}(4n^{3}+6n^{2}+8n+3)\\
E_{5}={
- {{\lambda }^2} \, {2^{-8}}
\left(
1572\,{n^5}+3930\,{n^4}+ 12220\,{n^3}
\right. }
\\
{\quad \quad
\left.
+ 14400\,{n^2}+ 11528\,n+3495
\right) }   \\
E_{7}={
5\,{{\lambda }^3} \,{2^{-11}}
\left(
23592\,{n^7}+82572\,{n^6}
\right. } \\
{ \quad \quad
\left.
+ 418236\,{n^5} + 839160\,{n^4}+ 1523968\,{n^3}
\right. } \\
{ \quad \quad
\left.
+ 1488078\,{n^2} + 939884\,n + 247935
\right) } \\
 E_{9}={
 - {{\lambda }^4} \, {2^{-16}}
 \left( 45804660\,{n^9}  + 206120970\,{n^8}
\right. } \\
{ \quad \quad
\left.
+ 1471569960\,{n^7}+ 4188597000\,{n^6}
\right. } \\
{ \quad \quad
\left.
+ 12317818548\,{n^5}+ 20804002800\,{n^4}
\right. } \\
{ \quad \quad
\left.
 +  29394281120\,{n^3}+ 25244303400\,{n^2}
 \right. } \\
{ \quad \quad
\left.
   + 13898196592\,n  +3342323355
 \right)
 } \\
 E_{11}= {
  5\,{{\lambda }^5} \,{2^{-19}}
 \left(
  1023655464\,{n^{11}}+ 5630105052\,{n^{10}}
\right. } \\
{ \quad \quad
\left.
+ 52379661180\,{n^9}+ 193482687420\,{n^8}
\right. } \\
{ \quad \quad
\left.
+ 801071289576\,{n^7}+ 1940241040920\,{n^6}
\right. } \\
{ \quad \quad
\left.
+ 4424265058200\,{n^5}+ 6633369121920\,{n^4}
\right. } \\
{ \quad \quad
\left.
+  8108461519360\,{n^3} + 6378900376878\,{n^2}
\right. } \\
{ \quad \quad
\left.
+ 3214574914460\,n + 725076383025
 \right) .
 }
}
\end{equation}

In conclusion, a new useful technique for deriving results of
the logarithmic
perturbation theory has been developed. Based upon the
$\hbar$ -expansions and suitable quantization conditions,
new handy recursion relations for solving the bound-state
 problem for an anharmonic
oscillator within the framework of the one-dimensional
Schr{\"o}dinger equation have
been obtained. Avoiding the disadvantages of the standard
approach these
formulae have the same simple form both for ground and
exited states and
provide, in principle, the calculation of the perturbation
corrections up to an
arbitrary order in the analytical or numerical form.
The extention on the three-dimensional
case and the
relativistic equations will be published elsewhere.
\ack
This work was supported in part by the International Soros
Science Education
Program (ISSEP) under Grant APU052102 .

\end{document}